\documentclass[conference]{IEEEtran}
\IEEEoverridecommandlockouts
\usepackage{cite}
\usepackage{amsmath,amssymb,amsfonts}
\usepackage{algorithmic}
\usepackage{graphicx}
\usepackage{textcomp}
\usepackage{xcolor}
\usepackage{multirow}
\usepackage{pdfpages}
\usepackage{balance}
\usepackage{array}
\newcolumntype{P}[1]{>{\centering\arraybackslash}p{#1}}
\def\BibTeX{{\rm B\kern-.05em{\sc i\kern-.025em b}\kern-.08em
    T\kern-.1667em\lower.7ex\hbox{E}\kern-.125emX}}
\begin{document}

\title{Enhanced U-Net: A Feature Enhancement Network for Polyp Segmentation\\
}

\author{\IEEEauthorblockN{Krushi Patel$^{\dag}$, 
Andrés M. Bur$^{\ddag}$, Guanghui Wang$^{*}$}
\IEEEauthorblockA{$^\dag$ \textit{Department of Electrical Engineering and Computer Science, University of Kansas, Lawrence KS, USA, 66045}\\
$^\ddag$ Department of Otolaryngology–Head and Neck Surgery, University of Kansas, Kansas City, Kansas, USA, 66160\\
$^*$ \textit{Department of Computer Science, Ryerson University, Toronto ON, Canada, M5B 2K3}\\
Corresponding author: wangcs@ryerson.ca
}
}

\maketitle

\begin{abstract}
Colonoscopy is a procedure to detect colorectal polyps which are the primary cause for developing colorectal cancer. However, polyp segmentation is a challenging task due to the diverse shape, size, color, and texture of polyps, shuttle difference between polyp and its background, as well as low contrast of the colonoscopic images. To address these challenges, we propose a feature enhancement network for accurate polyp segmentation in colonoscopy images. Specifically, the proposed network enhances the semantic information using the novel Semantic Feature Enhance Module (SFEM). Furthermore, instead of directly adding encoder features to the respective decoder layer, we introduce an Adaptive Global Context Module (AGCM), which focuses only on the encoder's significant and hard fine-grained features. The integration of these two modules improves the quality of features layer by layer, which in turn enhances the final feature representation. The proposed approach is evaluated on five colonoscopy datasets and demonstrates superior performance compared to other state-of-the-art models.

\end{abstract}


\section{Introduction}
Colorectal cancer is the third most common cancer diagnosed in the United States~\cite{states}. It is considered the second deadliest cancer in terms of mortality, causing 9.4\% of total cancer deaths~\cite{mort}. The primary reason behind colorectal cancer is a polyp that grows in the lining of the colon or rectum. Early detection and localization of polyp can reduce the mortality rate caused by colorectal cancer. In addition, it could reduce the treatment cost by restricting cancer spread to distant organs and the need for biopsy. Colonoscopy is the standard visual examination for the screening of colorectal cancer. However, the result of colonoscopy may be misleading due to the variant nature of polyps, including their shape, size, texture, and unpredictable factors such as veins and illumination. In addition, the result of colonoscopy depends on various human factors including gastrologist's experience  and physical and mental fatigue. Therefore, an automatic computer-aided polyp segmentation system is required to assist the physician during the procedure and significantly improve the polyp detection rate.

\begin{figure*}[t]
\begin{center}
\includegraphics[width = 0.8\textwidth]{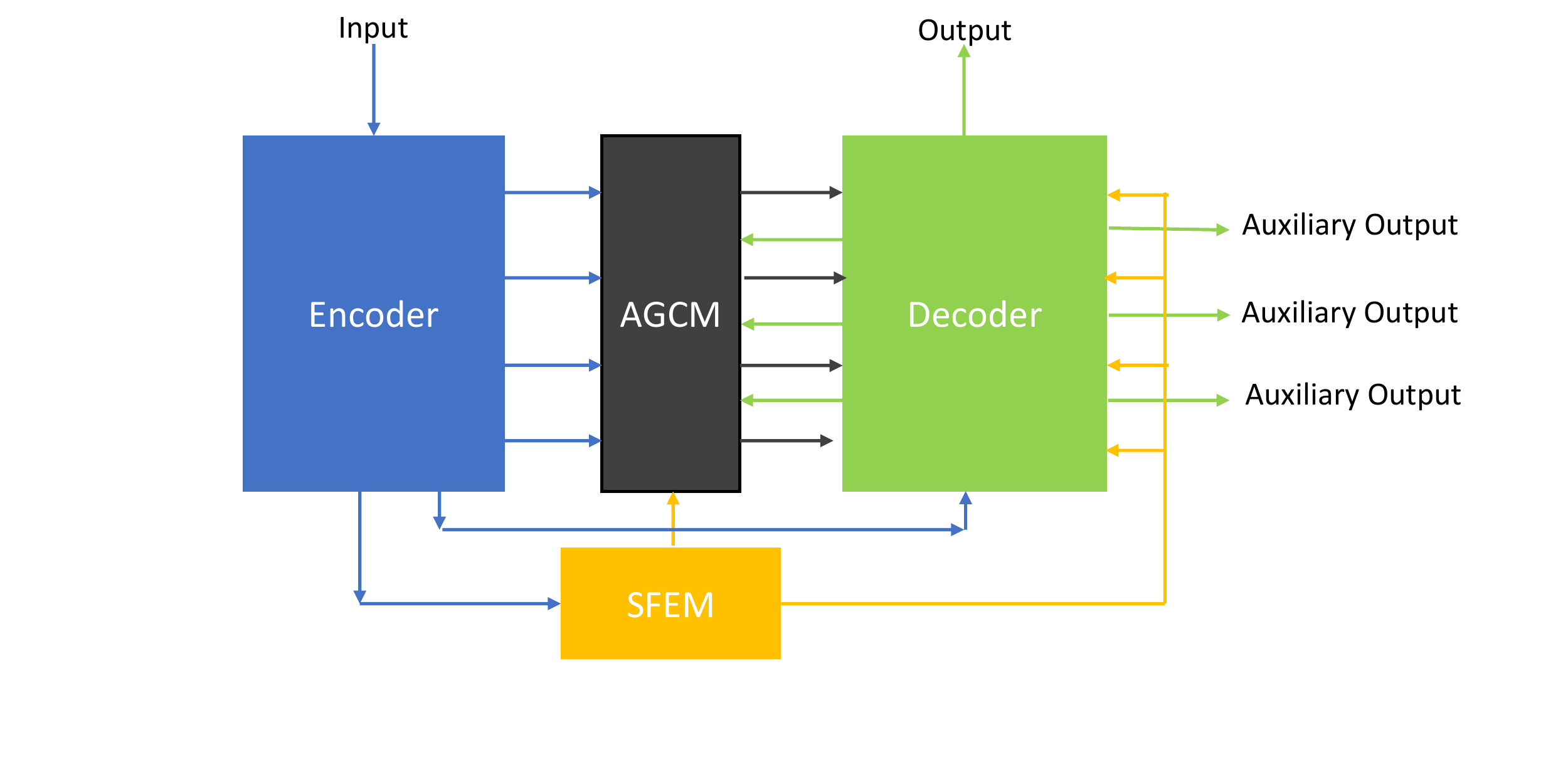}
\caption{The overall architecture of the enhanced U-Net: The input image is supplied to the encoder. Each encoder layer's features are sent to the respective decoder layer through the AGCM module. The features of the last encoding layer are applied to SFEM to further enhance the multi-scale semantic features.  The resultant features are sent to all decoder layers to concatenate with the features produced by AGCM and each decoder layer. Auxiliary losses are applied at the end of each decoder layer.} \label{fig:fig1}
\end{center}
\end{figure*}

Various techniques have been developed for the polyp segmentation task. The available methods can be largely divided into two categories:  (1) Hand-crafted feature based approaches and (2) Deep-learning based approaches \cite{li21}\cite{patel20}\cite{mo18}. Before the invention of neural networks, the polyp segmentation task depends on hand-crafted features such as size, shape, texture, and color~\cite{handfeature1}~\cite{handfeature2}.  However, these approaches are slow and have a high misdetection rate due to the low representation capability of hand-crafted features.  Following the huge success of deep learning-based models on generic datasets \cite{cen20}\cite{wu21}\cite{xu20}\cite{zhang20}, researchers started using neural networks for polyp detection and segmentation. Inspired by the early work~\cite{cnnfeature1},  where FCN~\cite{fcn} is utilized with a pre-trained model to segment the polyp, Akbari et al. \cite{cnnfeature2}  proposed a modified version of FCN to improve the performance of polyp segmentation. U-Net++~\cite{unet++}  and ResUNet++~\cite{resunet++} upgraded the architecture of U-Net~\cite{unet}  and achieved promising results on polyp segmentation. SFANet~\cite{sfa}  takes the area-boundary constraint into account along with extra edge supervision. It achieves good results but lacks generalization capability.  Recently introduced ACSNet~\cite{acs}  and PraNet~\cite{pra}  propose an attention-based mechanism to focus more on the hard region, which leads to improved performance.  

U-Net and its variants U-Net++, ResUNet, ResUNet++, and ACSNet have achieved appealing results on the polyp segmentation task by using U-shape encoder-decoder architecture. However, none of them utilize decoder features to calculate the attention value of the respective encoder layer. We believe that utilizing the decoder layer feature to selectively aggregate respective encoder layer features could improve the feature quality. Moreover, all of the above-mentioned models employ the pooling-based approach to enhance the multi-scale semantic features, which may lead to loss of spatial information. 

To alleviate these issues, we propose an attention-based U-Net for polyp segmentation by enhancing the quality of features. The proposed network mainly consists of two modules. First, we design a Semantic Feature Enhancement Module (SFEM), which enhances the deeper layer features by applying different sizes of patch-wise non-local attention block to tackle the different sizes of the polyp and fuse the output of each non-local blocks together. These fused features are then sent to each decoder layer.  Second, we introduce an Adaptive Global Context Module (AGCM), which focuses on more significant features of the encoder layer by taking into account the previous decoder layer features.  This mechanism suppresses the insignificant and noisy features and focuses only on essential features using spatial cross attention. It refines the decoder features layer by layer by removing unwanted features and adding significant fine-grained features only. Furthermore, to give more attention to the hard regions, we apply focal loss at each decoder layer.

In summary, the main contributions of the paper include: 

\begin{itemize}
\item The proposed semantic feature enhancement module fully exploits the multi-scale semantic context without losing spatial information. 
\item The proposed adaptive global context module attends the significant and hard fine-grained features and selectively aggregated them to the respective decoder layer. 
\item The integration of both modules enhances the quality of features layer by layer and hence achieves state-of-the-art performance on five publicly available benchmark datasets. 
\end{itemize}

The source code of the proposed model can be accessed at \url{https://github.com/rucv/Enhanced-U-Net}.

\section{Method}

\begin{figure*}[t]
\begin{center}
\includegraphics[width=1.0\textwidth]{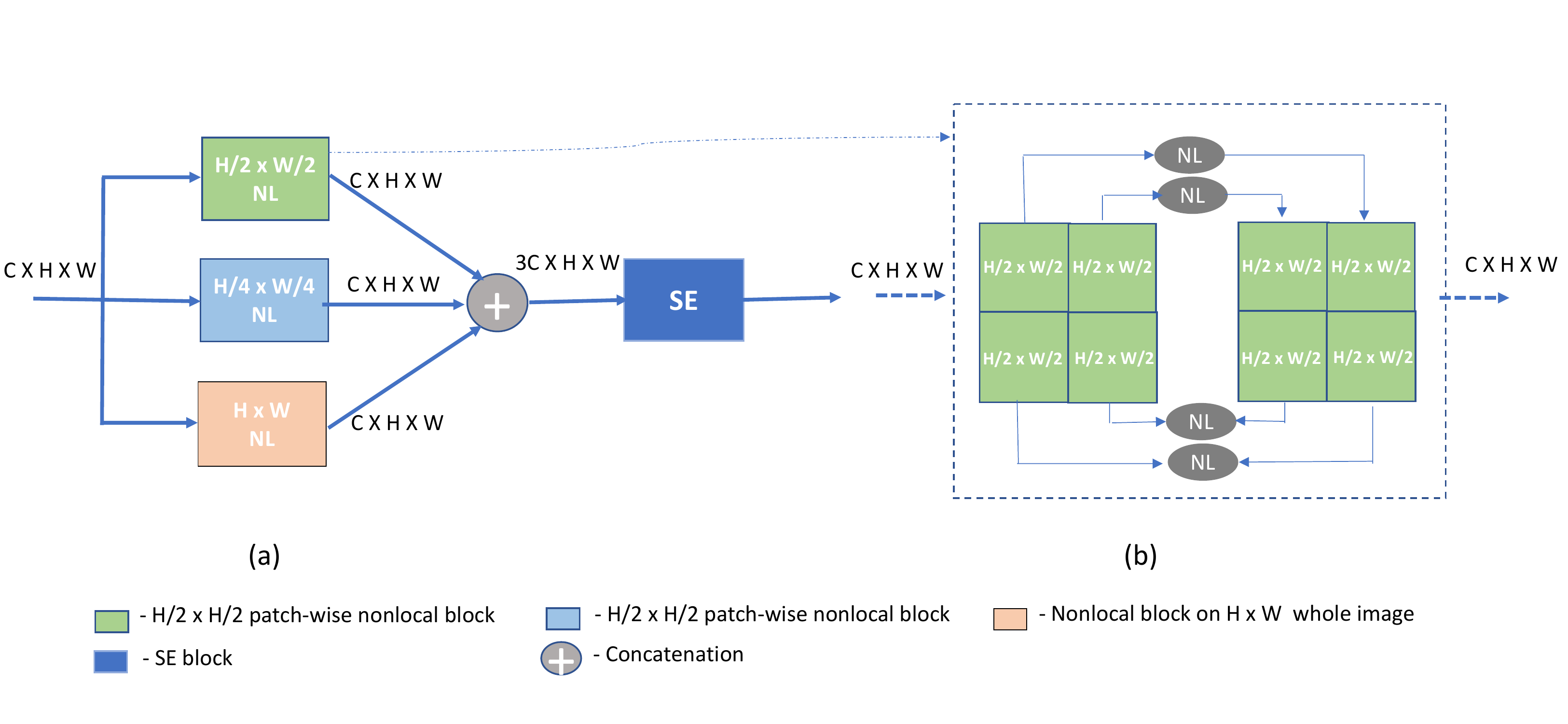}
\caption{(a) The detailed architecture of SFEM. It consists of three branches: the 1st and the 2nd branches divide the image into $H/2 \times H/2$ and $H/4 \times H/4$ sizes of patches and apply non-local attention. The third branch applies basic non-local attention to the whole image.  The detailed version of the non-local block on an individual patch is described in (b) where it first divides the image into patches and then applies non-local attention on each patch independently and folds it back to the whole image. The result of each branch is concatenated, followed by a SE-block.  } 
\label{fig:fig2}
\end{center}
\end{figure*}

\begin{figure}[!tb]
\begin{center}
\includegraphics[width=0.50\textwidth]{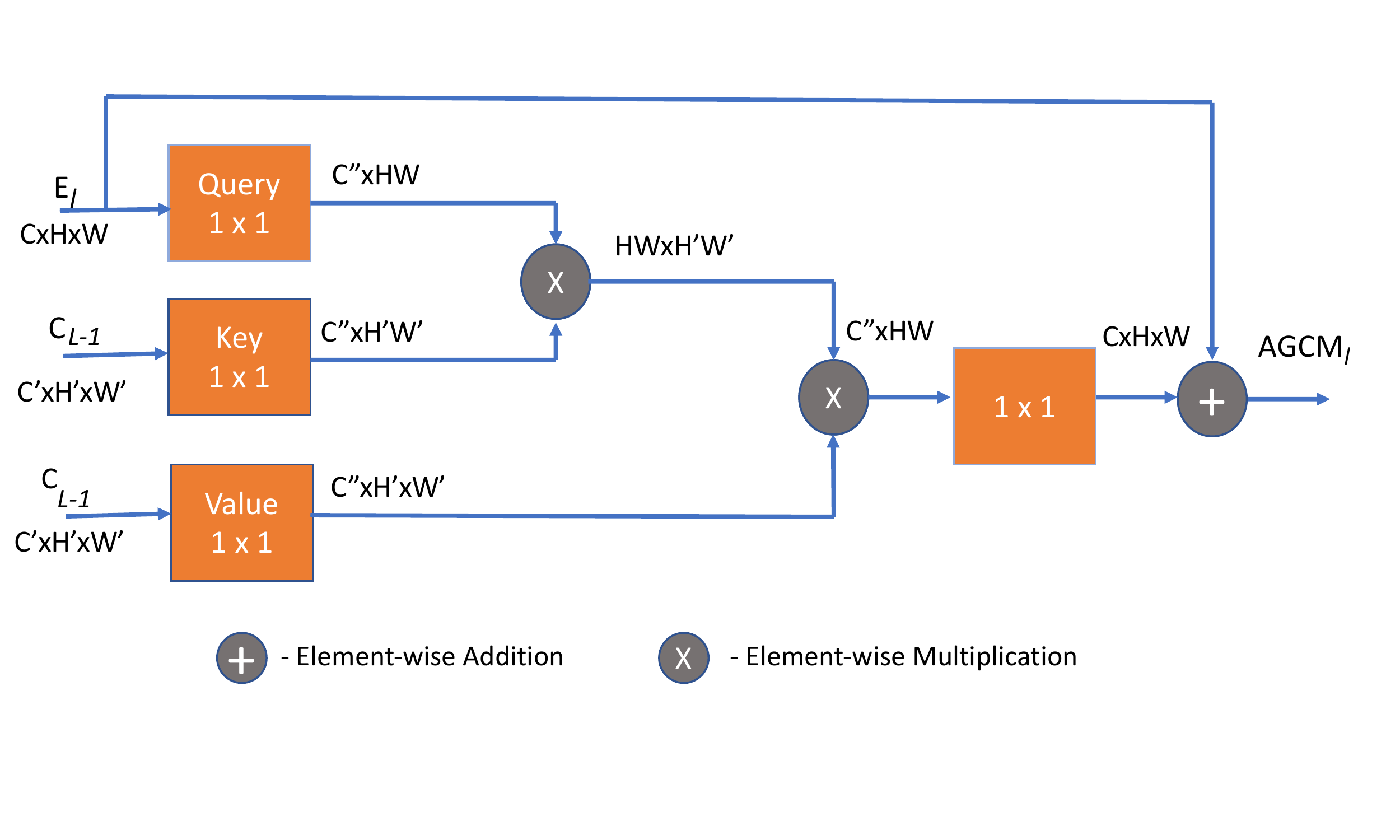}
\caption{The overall architecture of AGCM. It takes current encoder features $E_l$ as Query and concatenates features ($C_{l-1}$) generated from $(D_{l-1}, AGCM_{l-1} , SFEM)$ as Key and Value to perform cross attention.} \label{fig:fig3}
\end{center}
\end{figure}

The architecture of the proposed enhanced U-Net is shown in Fig~\ref{fig:fig1}. It mainly consists of four parts: (1) Encoder, (2) Decoder, (3) SFEM, and (4) AGCM. An encoder is made up of ResNet-34~\cite{res}. The encoder's output is fed to the decoder, which consists of five decoding layers. Each decoding layer consists of two convolution layers followed by batch-normalization and ReLU activation. The SFEM module is attached at the top of the last encoding layer, which consists of semantic features.  We insert one convolution layer before the SFEM module to reduce the number of channels. The output of SFEM is sent to all decoding layers. The AGCM module is employed in place of skip connection to alleviate the effect of background noise. It takes the current encoding layer and the previous decoding layer's feature maps as input and yields the resultant feature map of the same size as the current encoding layer feature map. Feature maps produce by SFEM, AGCM, and the decoder layer are concatenated and applied to the next decoding layer and AGCM. Each decoding layer is attached to the auxiliary loss inspired by deep supervision. 
The detailed description of the two proposed modules are as follows:

\subsection{Semantic Feature Enhancement Module}

It is well known that the deeper layers in CNN networks contain the semantic features which are most significant to detect and segment the objects. To fully exploit the semantic features,  we introduce a semantic feature enhancement module (SFEM) inspired by the pyramid pooling~\cite{pypool1}~\cite{pypool2}~\cite{pypool3}.

Specifically,  SFEM consists of three parallel branches of patch-wise non-local blocks as shown in Fig~\ref{fig:fig2}. It takes the output of the encoder feature map as input and applies non-local attention to the patches of a specific window size separately instead of applying adaptive average pooling. The first branch divides the image into four patches of size $(W/2 \times H/2)$, applies non-local spatial attention individually on each patch, and folded them back together as shown in Fig~\ref{fig:fig2}(b). Similarly, the second branch produces 16 patches of the size $(W/4 \times H/4)$ and performs the same operation as the first branch on each patch. In our experiment, we set the size of the output feature map of the encoder to $8 \times 8$. Therefore, the first branch contains the 4 patches of size $4\times 4$, and the second branch has the 16 patches of size $2 \times 2$. The last branch performs a non-local\cite{nonlocal} operation on the entire feature map of size $8 \times 8$. The outputs of these three branches are concatenated, followed by a squeeze and excitation block that attends to the most significant channels. The results of SE blocks\cite{se} are then sent to all decoder layers. To match each decoder layer's size, we upsample the output of SFEM. 

Unlike pyramid pooling, the above SFEM module is capable of enhancing the semantic information without losing spatial information. In SFEM, the size of each branch's output is the same, whereas for pyramid pooling, as the window size increases, the output size decreases, which requires an upsampling operation that leads to loss of spatial information.

\subsection{Adaptive Global Context Module}

Features generated using the SFEM module are at a coarse level and contain noise in it.  We propose an adaptive global context module (AGCM) to improve these coarse level features to fine level features layer by layer using spatial cross-layer attention.  The detailed architecture of the AGCM module is shown in Fig~\ref{fig:fig3}. It takes the current encoder feature map as query and concatenated features of SFEM, previous layer AGCM, and decoder layer as a key and value pair and applies cross-layer spatial attention\cite{attn}. The resultant attention features have the same size as the encoder layer feature map, so they can be directly aggregated to the encoder feature map without resizing operation. The aggregated features are then sent to the respective decoder layer. A detailed explanation has been given below.

In context to our encoder-decoder architecture, the basic non-local block can be formulated as:


\begin{equation}
y_{i} = \frac{1}{C(e_{l})} \sum_{\forall j} f(e_{il}, e_{jl}) g(e_{jl})
\end{equation}
where  $e_{l}$ is the features of  the encoder layer. $l$.  $e_{il} and e_{jl} $ come from the same encoder layer $l$  and produce the relationship matrix of size $WH \times WH$, where W and H denote the width and height of the encoder feature map.  In contrast, our AGCM can be formulated as:

\begin{equation}
y_{i} = \frac{1}{C(e_{l})} \sum_{\forall j} f(e_{il},  c_{j(l-1)}) g(c_{j(l-1)})
\end{equation}

Here, instead of using the same encoding layer features,  we established the global relationship between the current encoding layer $e_l$ and the fusion features $c_{l-1}$ generated by previous decoder layer features$d_{l-1}$, previous AGCM features $AGCM_{l-1}$ and SFEM features $SFEM_{l}$. This mechanism leads to selectively aggregating  fine-grained features of each encoding layer to the decoding layer instead of directly aggregating using addition or concatenation.

\begin{table*}[!htb]
\begin{center}
\caption{Segmentation masks produced by our model. The first column consists of the original image. The second column represents the ground truth mask and the third column shows the output mask generated by our model}\label{table:tab6}
\begin{tabular}{P{3.5cm}P{3.5cm}P{3.5cm}}

 Original  &Ground-Truth&Output\\

\includegraphics[width=\linewidth, height=40mm]{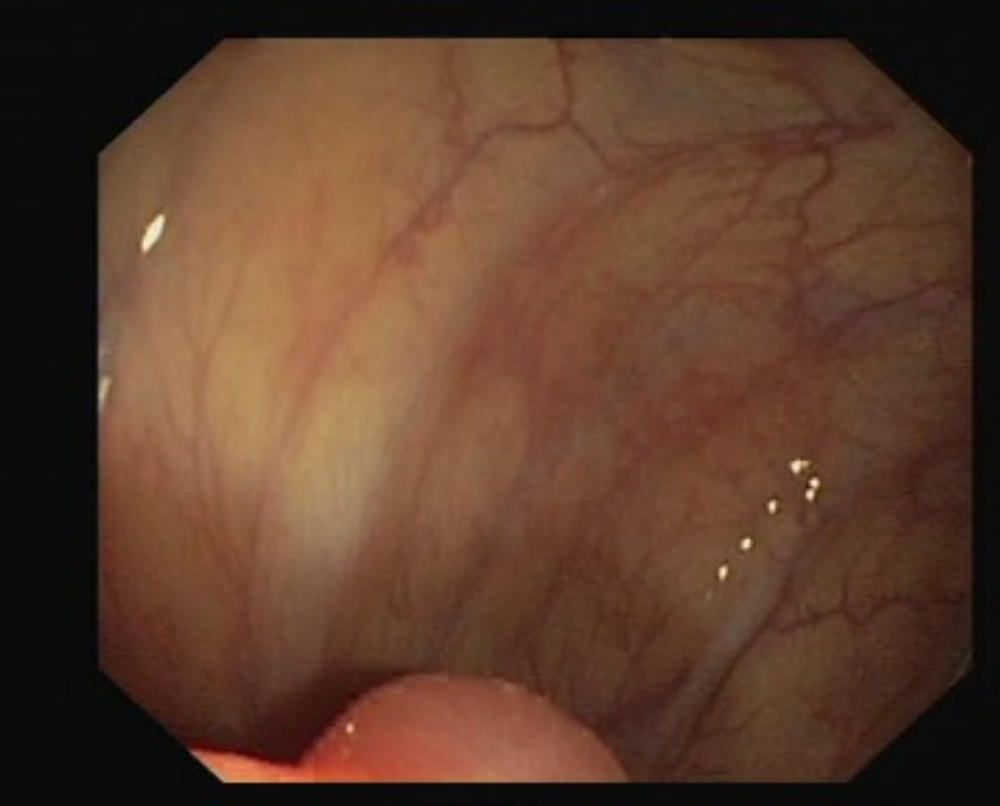}&\includegraphics[width=\linewidth, height=40mm]{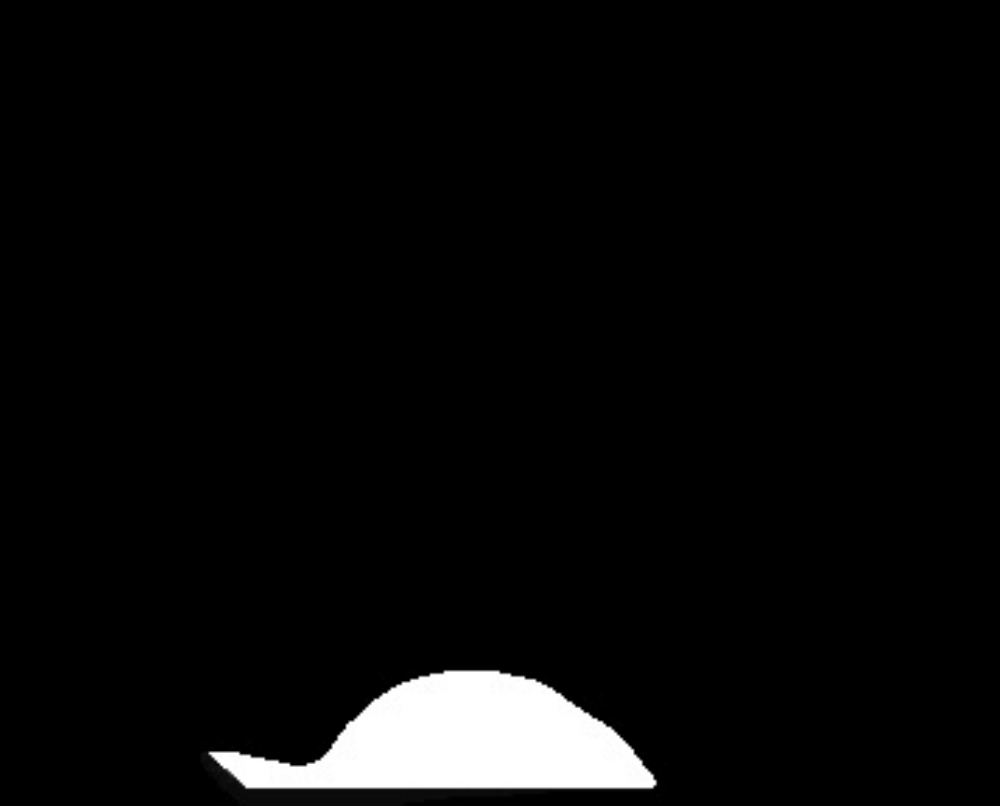} &
 \includegraphics[width=\linewidth, height=40mm]{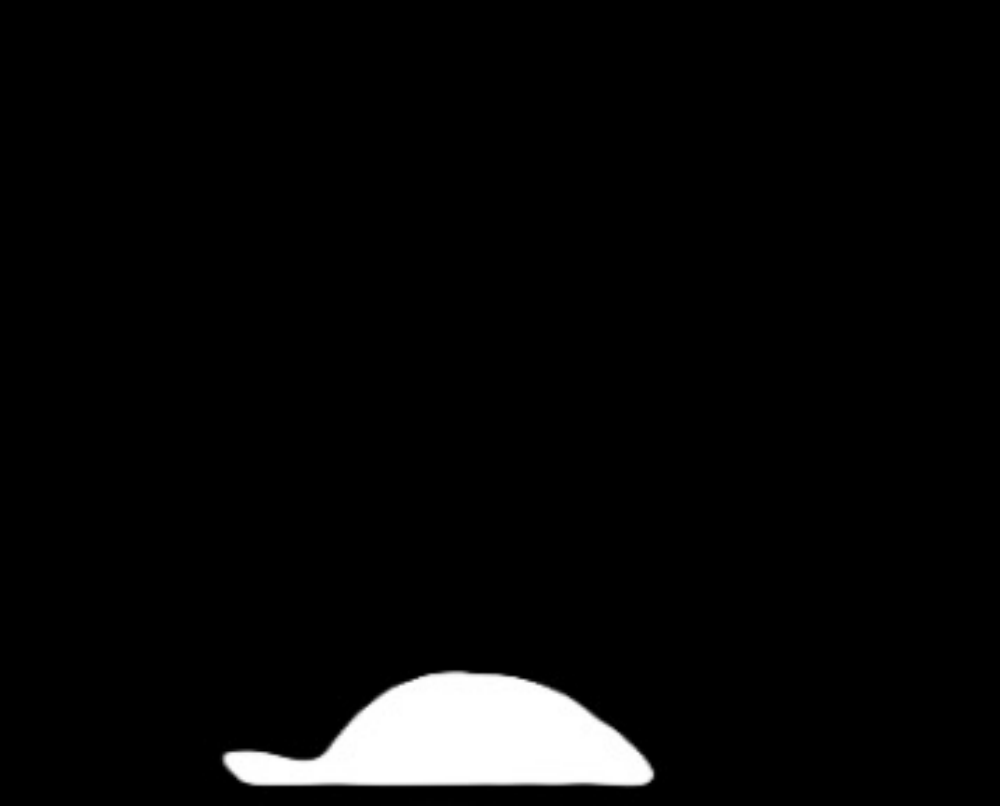}\\
 
\includegraphics[width=\linewidth, height=40mm]{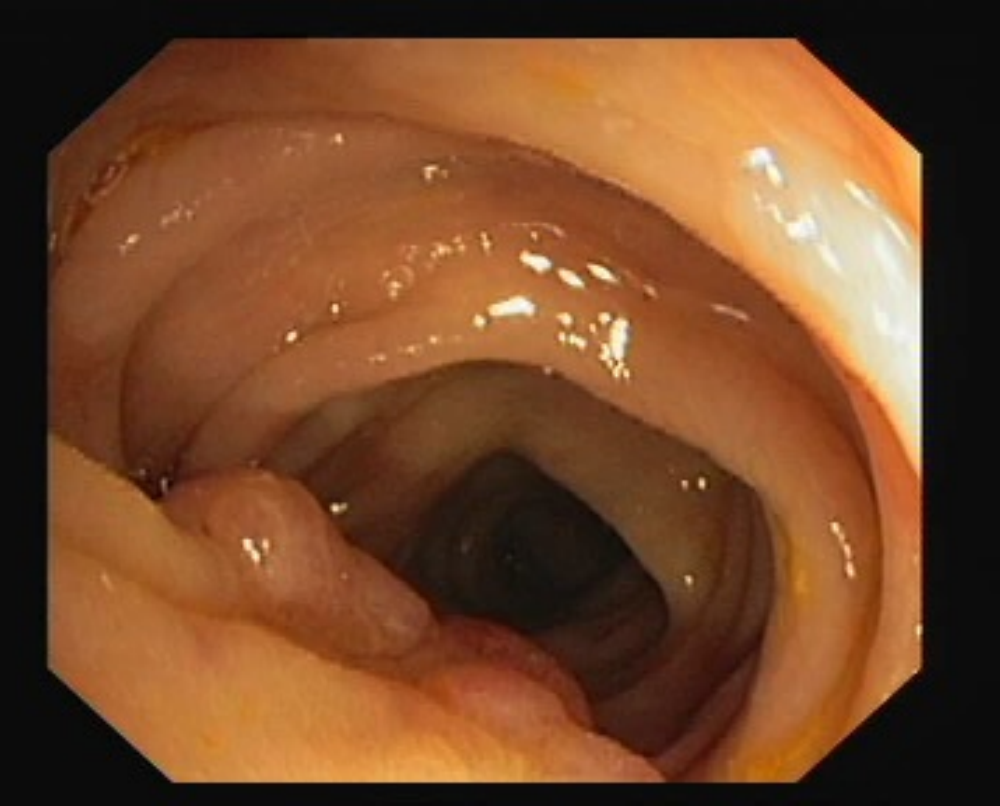}&\includegraphics[width=\linewidth, height=40mm]{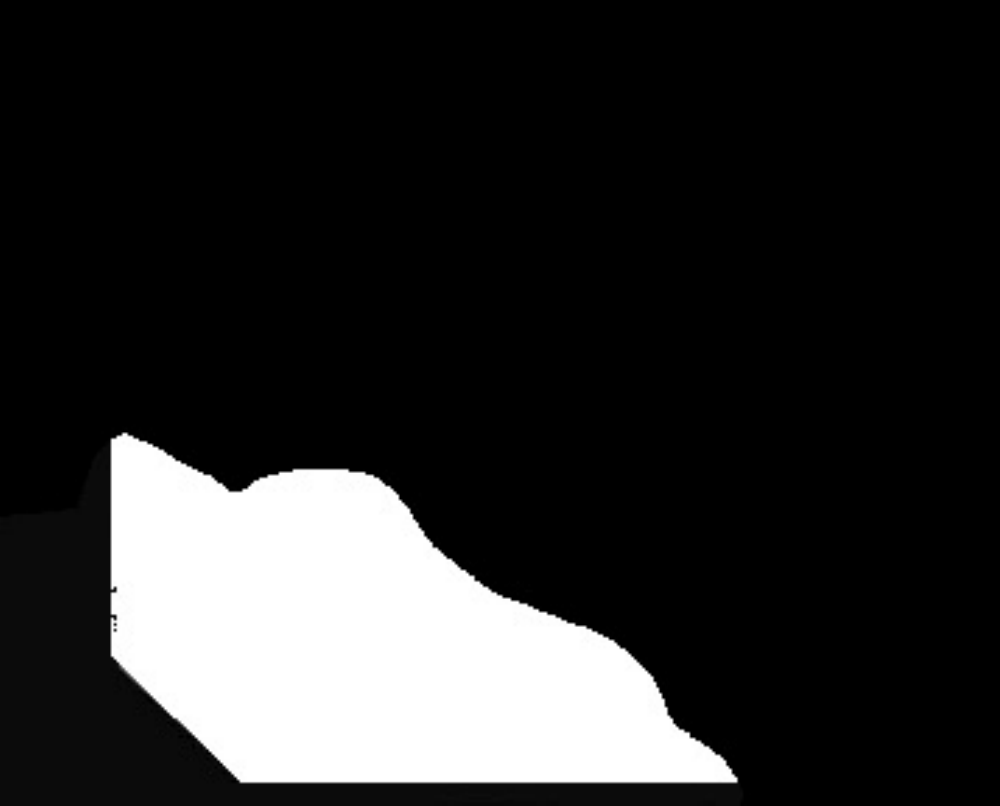} &
 \includegraphics[width=\linewidth, height=40mm]{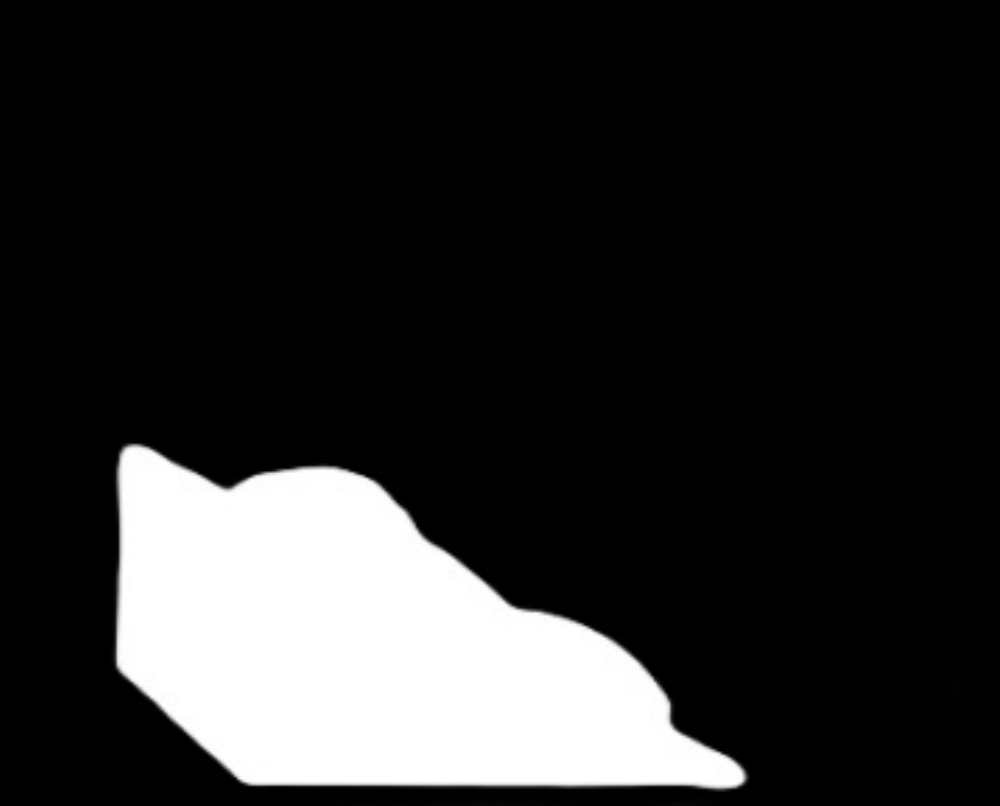}\\
 
\includegraphics[width=\linewidth, height=40mm]{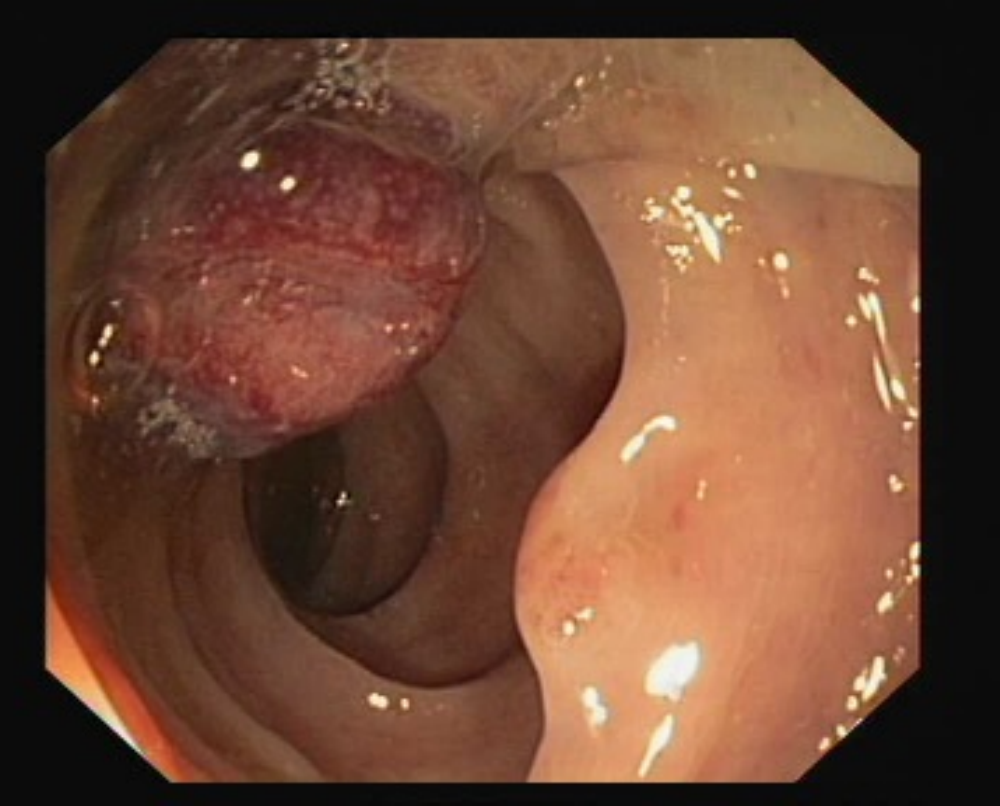}&\includegraphics[width=\linewidth, height=40mm]{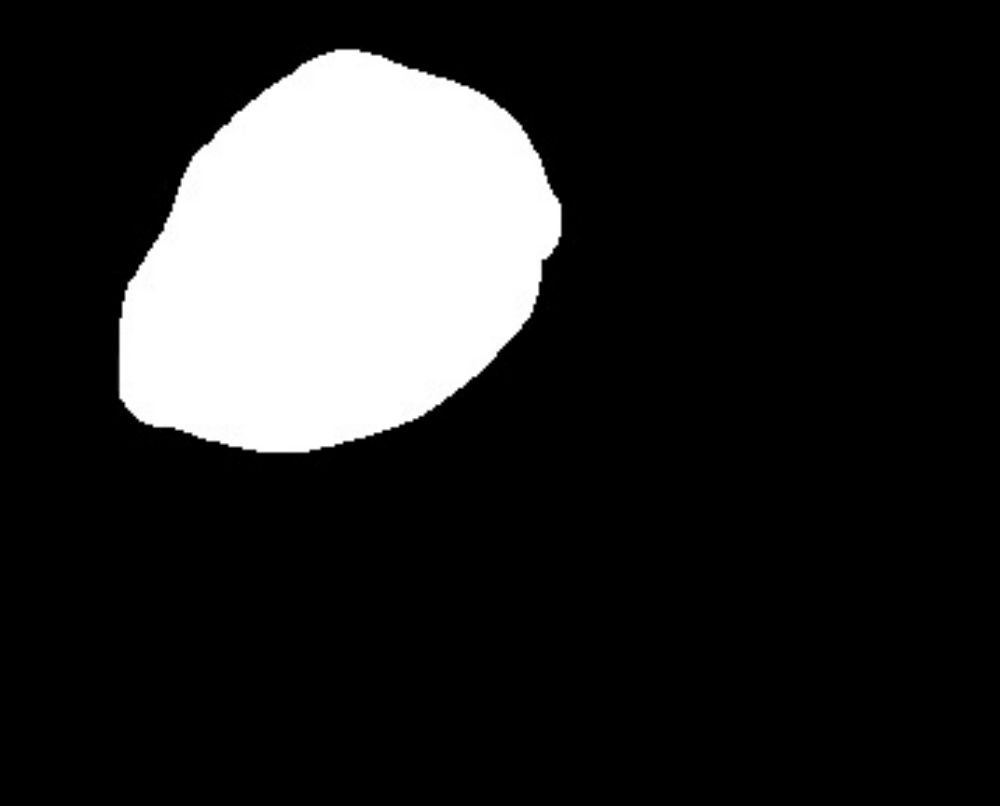} &
 \includegraphics[width=\linewidth, height=40mm]{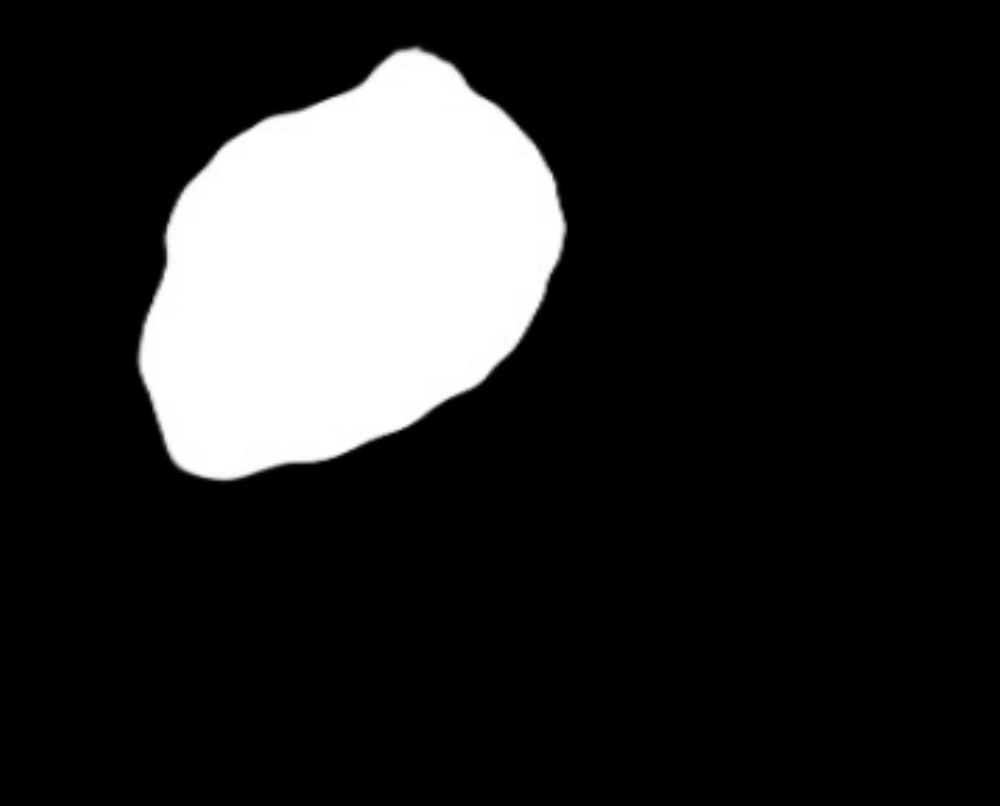}\\
 
\includegraphics[width=\linewidth, height=40mm]{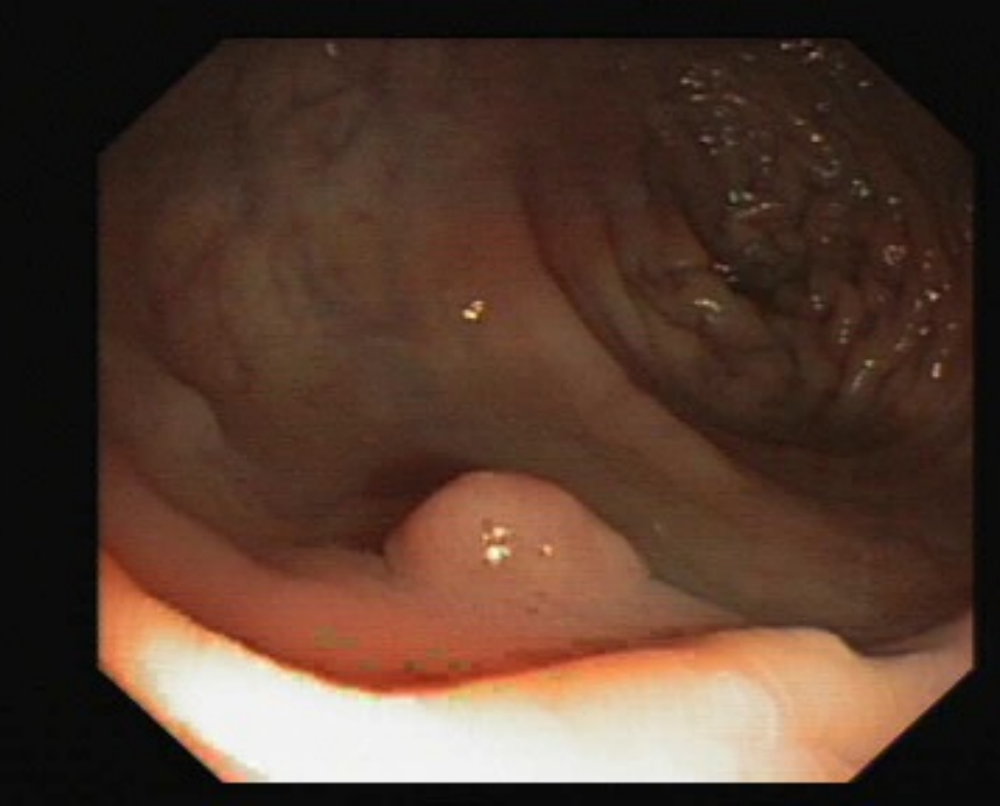}&\includegraphics[width=\linewidth, height=40mm]{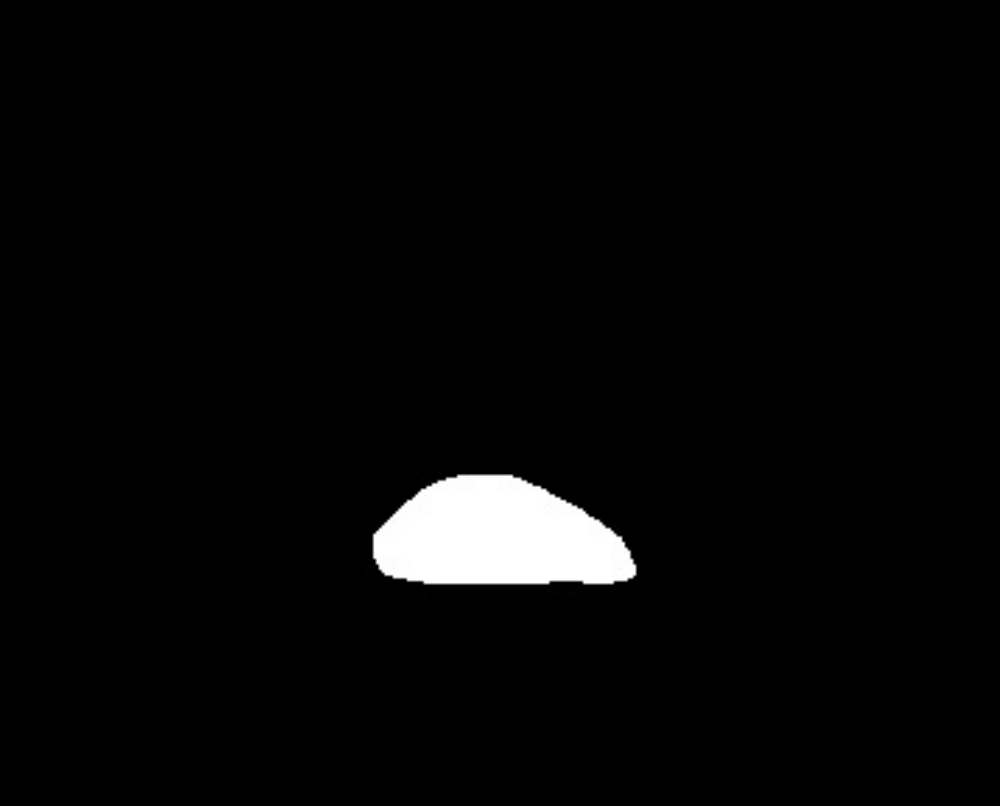} &
 \includegraphics[width=\linewidth, height=40mm]{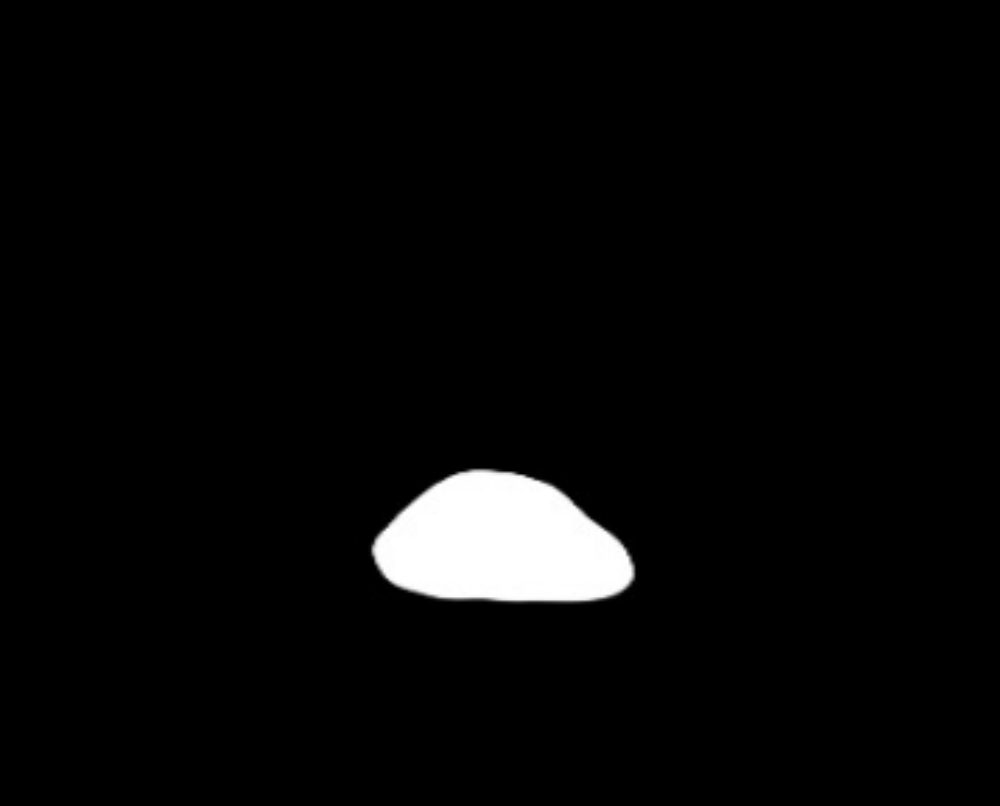}\\
 

\end{tabular}
\label{tab6}
\end{center}
\end{table*}

\begin{table*}[!htb]
\begin{center}
\caption{Results on the EndoScene dataset to prove the learning capability. Train and Test sets are from the same dataset.}\label{table:tab1}
\begin{tabular}{l|p{1.5cm}|p{1.5cm}|p{1.5cm}|p{1.5cm}|p{1.5cm}|p{1.5cm}}
\hline
Models&  Recall & Specificity&Precision&Dice&IoU&Accuracy\\
\hline
\hline
FCN8 & 60.21 &98.60&79.59&61.23&48.38&93.77\\
UNet&85.54&98.75&83.56&80.31&70.68&96.25\\
UNet++& 78.90 & 99.15&86.17&77.38&68.00&95.78\\
SegNet& 86.48&99.04&86.54&82.67&74.41&96.62\\
SFANet& 85.51&98.94&86.81&82.93&75.00&96.61\\
ACSNet&90.18&99.19&93.13&90.27&85.31&98.13\\
Ours&\textbf{92.55}&\textbf{99.47}&\textbf{93.54}&\textbf{92.40}&\textbf{86.74}&\textbf{98.97}\\
\hline
\end{tabular}
\end{center}
\end{table*}

\begin{table*}
\begin{center}
\caption{Results on the Kvasir-SEG dataset to prove the learning capability. Train and Test sets are from the same dataset.}\label{table:tab2}
\begin{tabular}{l|p{1.5cm}|p{1.5cm}|p{1.5cm}|p{1.5cm}|p{1.5cm}|p{1.5cm}}
\hline
Models&Recall& Specificity&Precision&Dice&IoU&Accuracy\\
\hline
\hline
UNet&87.89&97.69&83.89&82.85&73.95&95.65\\
UNet++& 88.67 & 97.49&83.17&82.80&73.74&94.49\\
ResUNet &81.25&98.31&87.88&81.14&72.23&94.90\\
SegNet& 90.03&98.13&87.15&86.43&79.11&96.68\\
SFANet& 91.99&97.05&82.95&84.68&77.06&95.71\\
ACSNet&93.76&\textbf{98.02}&91.94&92.23&87.20&\textbf{97.74}\\
Ours&\textbf{93.89}&97.92&\textbf{92.69}&\textbf{92.58}&\textbf{87.56}&97.69\\
\hline
\end{tabular}
\end{center}
\end{table*}

\begin{table*}[!htb]
\begin{center}
\caption{Results on ColonDB,  ETIS, and CVC-300 to prove the generalization capability of the model. Train and Test sets come from different datasets.}\label{table:tab3}
\begin{tabular}{p{2cm}|p{2cm}|p{2cm}|p{2cm}|p{2cm}}
\hline

Dataset&Models&mean Dice&mean IoU&Accuracy\\
\hline
\hline
\multirow{5}{4em}{ColonDB}&U-Net&51.2&44.4&93.9\\
&U-Net++&48.3&41.0&93.6\\
&SFA&46.9&34.7&90.6\\
&Pra-Net&70.9&64.0&\textbf{95.5}\\
&Ours&\textbf{73.99}&\textbf{66.28}&95.35\\
\hline
\multirow{5}{4em}{ETIS}&U-Net&39.8&33.5&96.4\\
&U-Net++&40.1&34.4&96.5\\
&SFA&29.7&21.7&89.1\\
&Pra-Net&62.8&56.7&\textbf{96.9}\\
&Ours&\textbf{65.07}&\textbf{58.20}&96.50\\
\hline
\multirow{5}{4em}{CVC-300}&U-Net&71.0&62.7&97.8\\
&U-Net++&70.7&62.4&98.2\\
&SFA&46.7&32.9&93.5\\
&Pra-Net&87.1&79.7&99.0\\
&Ours&\textbf{88.62}&\textbf{81.30}&\textbf{99.26}\\
\hline

\end{tabular}
\end{center}
\end{table*}

\subsection{Loss Function}
Our loss function is defined as:
\begin{equation}
L = L_{IoU} + L_{Focal} + L_{Dice}
\end{equation}
where, $L_{IoU}$, $ L_{Focal}$ and $L_{Dice}$ represent the pixel-based IoU loss, focal loss and dice loss\cite{focal}\cite{dice}\cite{iou}\cite{wiou}. To take the individual benefits of each loss function, we have combined the above three losses. Focal loss is employed to put more focus on the difficult pixels concerning probability score. In addition, we have used the weighted IoU loss to give more weight to the harder pixels based on their neighborhood pixels. At last, to focus on the foreground object, we use dice loss. Our experiment shows that adding dice loss with weighted IoU loss increases the performance by a significant amount. In addition, we utilize deep supervision for all decoder layer prediction maps generated by side-out. We downsample the ground truth mask to match the size of the prediction generated by the appropriate decoding layer.

\begin{table*}[!htb]
\begin{center}
\caption{Ablation study on set-1 and set-2.  Performance comparison of the model using only SFEM, only AGCM, or both together.}\label{table:tab4}
\begin{tabular}{|l|p{1.5cm}|p{1.5cm}|p{1.5cm}||p{1.5cm}|p{1.5cm}|p{1.5cm}|}
\hline
 \multirow{3}{4em}{Models}&\multicolumn{3}{|c|}{Set-1}& \multicolumn{3}{|c|}{Set-2} \\
\hline
&Mean Dice& Mean IoU & Accuracy&Mean Dice& Mean IoU & Accuracy\\
\hline
\hline
Baseline(U-Net)&80.31&70.68&96.25&71.0&62.7&97.8\\
Baseline+ fcloss + IoUloss + Diceloss&88.77&83.81&98.33&78.52&70.27&97.37\\
Baseline + our loss + SFEM&91.39&85.46&\textbf{99.04}&83.56&76.83&98.08\\
Baseline + our loss +AGCM&91.31&85.34&\textbf{99.04}&83.13&76.33&98.13\\
Baseline + our loss + SFEM +  AGCM&\textbf{92.40}&\textbf{86.74}&98.97&\textbf{88.62}&\textbf{81.30}&\textbf{99.26}\\
\hline
\end{tabular}
\end{center}
\end{table*}

\section{Experiments}

\subsection{Datasets}

We evaluate the proposed model on the following five benchmark datasets for polyp segmentation: ETIS \cite{etis}, CVC-ClinicDB \cite{col1},  CVC-ColonDB \cite{col2},  Endoscene \cite{endo}, and Kvasir \cite{kavsir}. ETIS is an old dataset that consists of 196 polyp images. CVC-ClinicDB contains 612 polyp images from 29 colonoscopy videos. EndoScence combines the CVC-ClinicDB and CVC-300 dataset, where CVC-300 consists of 300 images from 13 short colonoscopy sequences. CVC-ColonDB is a small-scale database that has 380 images from 15 short coloscopy sequences. At last, Kvasir is a recently proposed challenging dataset that consists of 1000 images with its ground-truth masks. 
We compare the enhanced U-Net with the baseline models: U-Net~\cite{unet}, U-Net++~\cite{unet++}, and ResUNet++~\cite{resunet++}.  We also compare the performance of our model with the recently proposed ACS~\cite{acs} and PraNET~\cite{pra}.  Specifically, we perform the experiments in two modes of the dataset: Set-1 and Set-2.  For the first mode, Set-1, we divide the Kavasir-SEG and CVC-ColonDB datasets into Train, Val, and Test set individually. In contrast, for the second mode, Set-2, we combine both datasets and used them to train the model and evaluate performance on a totally different dataset, including ETIS, CVC-300, and CVC-ColonDB.  

\subsection{Implementation Details}

During training, we resize all images of the Kavasir-SEG dataset to 384 X 288 and the remaining dataset to 320 X 320 and then randomly crop the images of size 256 X 256. We utilize several data augmentation methods to reduce the overfitting, including horizontal and vertical flips, rotation, and zoom. We set the batch size to 4 and train the model for 150 epochs with an initial learning rate of 0.001. We employ the SGD optimizer with a momentum of 0.9 and weight decay of 0.0005. 

To evaluate the permanence, we use recall, precision, specificity, dice-score, IoU, and accuracy as evaluation metrics. To make a fair comparison, we follow the same procedure to calculate the metric as ACM and PraNet.

\subsection{Results}

We compare the performance of our "Enhanced U-Net" with FCN\cite{fcn}, U-Net\cite{unet}, U-Net++\cite{unet++}, SegNet\cite{seg}, SFANet\cite{sfa}, and ACSNet\cite{acs} on Endoscene and the recently released Kvasir-SEG datasets. Table~\ref{table:tab1} and Table~\ref{table:tab2}  show the results on EndoScene and Kvasir-SEG datasets, respectively. Our model outperforms all the above state-of-the-art models with an adequate margin on almost all metrics.
Specifically, our model increases the Dice and IoU by 12.09\% and 16.06\% on the Endoscene dataset and 9.73\% and 13.61\% on the Kvasir dataset, respectively, compared to the baseline U-Net. It also outperforms the ACSNet by improving the majority of metrics 
by significant amount on both datasets. This indicates the effective learning ability of our model to segment the polyp.

To validate the generalization capability of our method, we further evaluate the performance of our model using new datasets that have never been seen before. We follow the same procedure to calculate the mean-IOU, mean Dice,  and Accuracy and utilize the same train and test set as PraNet\cite{pra} for a fair comparison. We then evaluate and compared the performance of different models using the following new datasets: ColonDB,  ETIS, and CVC-300. The results are shown in Table~\ref{table:tab3}. It is evident that our model improves the mean-Dice and mean-IOU by 22.79\% and 21.48\% on the ColonDB dataset, 25.25\% and 24.7\% on the ETIS dataset, and 17.62\% and 18.6\% on the CVC-300 dataset compare to the baseline U-Net. It also outperforms the recently proposed PraNet by increasing the mean-Dice and mean-IoU by an adequate amount. 
In short, it outperforms state-of-the-art methods on the majority of metrics with a significant margin, which demonstrates the superior generalization capability of the method.  

Furthermore, we also display the output of segmentation masks generated by our model in the Table~\ref{table:tab6}. From the result, it can be seen that ground truth masks and the outputs look very similar such that they are hard to differentiate. 

\subsection{Ablation Study}

In this section, we present the ablation experiments to validate the effectiveness of our proposed modules individually. We train U-Net baseline on both modes of dataset and test on EndoScene and CVC-300 datasets by either including SFEM, AGCM or proposed loss function as well as inlcluding all or two modules together. The results of the ablation study are shown in Table  ~\ref{table:tab4}.

\subsubsection{Effect of Losses}
We first validate the proposed composite loss function's effectiveness by adding only that loss into the baseline. It can be seen from Table ~\ref{table:tab4} that the proposed loss function with deep supervision tremendously improves the performance. Specifically, for set-1, mean Dice and mean IoU increase by 8.46\% and 13.13\% respectively, and for set-2, mean Dice and mean IoU improve by 7.52\% and 7.57\% respectively. From Table  ~\ref{table:tab4}, it can also be seen that adding dice loss along with focal loss and weighted IoU loss increase the performance by a significant amount. 

\subsubsection{Effect of SFEM}

From the result in Table ~\ref{table:tab4} it can be seen that only including SFEM improves the performance of the baseline network in both test sets. 
For the set-1 dataset, mean Dice, mean IoU, and Accuracy are increased by 2.62\%, 1.65\%, and 0.74\%, respectively, compared to the baseline U-Net, indicating the improvement of our model's learning ability. The SFEM module dramatically improves the performance on set-2. Specifically, the mean Dice, mean IoU, and Accuracy on the set-2 dataset increase significantly by 5.04\%,6.56\%, and 0.71\%, respectively, compared to the baseline network, which indicates the generalization capability of SFEM.


\subsubsection{Effect of AGCM:}
Like SFEM, AGCM also improves the performance on both set-1 and set-2 datasets compared to the baseline network shown in Table ~\ref{table:tab4}. Specifically,  for set-1, mean Dice,  mean Iou, and accuracy are improved by 2.54\%,  1.53\%, and 0.71\%. For set-2, mean IOU and mean Dice improves dramatically by 4.61\% and 6.06\%, respectively, and accuracy is increased by 0.76\%. The improvements prove the model's earning and generalization capability by introducing AGCM, compared to the baseline.

\section{Conclusion}

This paper has presented a novel architecture to improve the quality of features layer by layer for automatically polyp segmentation from colonoscopy images. Our extensive experiments prove that our model consistently outperforms the baseline network U-Net and its variants: U-Net++ and ResUNet, by a large margin on different datasets. It also outperforms the recently published ACSNet and PraNet by a significant margin. The experiments demonstrate the strong learning capability and generalization ability of the proposed model. The proposed model could also be directly applied to other medical image segmentation tasks.

\section*{Acknowledgement}
This work was partly supported by the National Institutes of Health (NIH) under grant no 1R03CA253212-01 and the Nvidia GPU grant.

\balance 

\balance 

\end{document}